\documentclass[showpacs,preprintnumbers]{revtex4}
\usepackage{amssymb}
\usepackage{graphicx}

\begin{document}

\title{Spin and orbital ferromagnetism in strongly correlated
itinerant-electron systems}
\author{V. Yu. Irkhin$^*$ and M. I. Katsnelson}

\affiliation{Institute of Metal Physics, 620219 Ekaterinburg,
Russia}

\affiliation{Institute for Molecules and Materials, Radboud
University Nijmegen, 6525 ED Nijmegen, The Netherlands}

\begin{abstract}
Spectra of one-particle and collective excitations in narrow-band
ferromagnets with unquenched orbital moments are calculated in various
theoretical models. The interaction of spin and orbital excitations with
conduction electrons results in the damping of the former which, however,
turns out to be rather small; therefore, apart from usual spin waves,
well-defined orbitons can exist. Non-quasiparticle states occur in the
electron energy spectrum near the Fermi energy due to this interaction. The
criteria of stability of the saturated spin and orbital ferromagnetic
ordering are considered. Possible effects of orbital ordering in magnetite
and in colossal magnetoresistance manganites are discussed.
\end{abstract}

\maketitle

\section{Introduction}

Orbital ordering in strongly correlated systems is a subject of
interest for a rather long period \cite{kugel}, this interest
growing considerably last time (see, e.g., recent papers
\cite{khali,nagaosa,saitoh,kikoin,jeroen}). For orbitally ordered
magnets the energy spectrum is more rich than for purely
spin-moment magnetic materials since it contains additional
excitation branches (``orbitons''). Relevance of the orbital
degrees of freedom for physics of the colossal magnetoresistance
(CMR) manganites \cite{nagaosa,saitoh,salamon,nagaev,allen} is one
of the reasons for the revival of this field. A large orbital
contribution to the total magnetic moment has been observed
recently by the X-ray magnetic circular dichroism (XMCD) method
for magnetite Fe$_3$O$_4$ \cite{huang}. The latter substance (the
most widespread natural iron compound) is probably one of the most
intriguing magnetic oxides because of controversial experimental
information concerning its ground state and so called ``Verwey
transition'' \cite {magnetite,subias} (see also a discussion in
Ref.\onlinecite{physlett}). The magnetite is also a promising
material: it is a half-metallic ferromagnet (more exactly,
ferrimagnet) with a high Curie temperature and perspectives of
applications in spintronics \cite{UFN}. Since current theoretical
description of the half-metallic ferromagnetism (for a review, see Refs.%
\onlinecite{UFN,condmat}) ignores the orbital degrees of freedom, it should
be generalized to include them and to take into account probable orbital
ferromagnetism in the magnetite and, possibly, in other strongly correlated
conducting ferromagnets. Such a generalization is the aim of the present
work.

The paper is organized as follows. In Section 2 we formulate the models of a
narrow-band ferromagnet including both spin and orbital degrees of freedom
and discuss their relevance for the colossal magnetoresistance (CMR)
manganites and for the magnetite. In Section 3 the Bose-type Green's
functions are calculated which determine the spectrum and damping of spin
waves (magnons) and orbitons. In Section 4 the Fermi-type one-particle
Green's functions are derived and peculiarities of the electron energy
spectrum owing to the electron-magnon and electron-orbiton interactions are
considered. Using the expressions obtained the conditions for stability of
saturated spin and orbital ferromagnetic ordering are investigated. In
Section 5 a general picture of the half-metallic ferromagnetism in the
presence of the orbital degrees of freedom is discussed.

\section{Models of a conducting ferromagnet with orbital degrees of freedom}

Previous works concerning the role of the orbital degrees of
freedom in strongly correlated magnetic systems treated mainly the
insulator case, in particular, orbital ferromagnetism accompanied
with spin antiferromagnetism \cite{kugel}, orbital
antiferromagnetism \cite{kikoin}, or orbital liquid \cite{khali}.
In Ref.\onlinecite{brink} a mean-field phase diagram of an
insulator including orbital and spin ferromagnetism region was
obtained. In contrast with these works, we consider here
\textit{conducting} ferromagnets with both spin and orbital
degrees of freedom.

We start with the many-electron system with two ground terms of
the $d^n$ and $d^{n+1}$ configurations, $\Gamma _n=\{SL{\}}$ and
$\Gamma_{n+1}=\{S^{\prime }L^{\prime }{\}}$. The corresponding
periodic Anderson model that describes the electron hopping with
the transitions between two these states only reads
\begin{equation}
\mathcal{H}=\mathcal{H}_0+\sum_{\mathbf{k}\sigma m}t_{\mathbf{k}m}c_{\mathbf{%
k}\sigma m}^{\dagger }c_{\mathbf{k}\sigma m}+\sum_{\mathbf{k}im\sigma
}\left( V_{\mathbf{k}}c_{\mathbf{k}\sigma m}^{\dagger }e^{i\mathbf{kR}%
_i}a_{il\sigma m}+\mathrm{h.c.}\right)  \label{m1}
\end{equation}
where $\mathcal{H}_0$ is the Hamiltonian for strongly correlated $d$%
-electrons described by the operators $a_{il\sigma m}$, $t_{\mathbf{k}m}$ is
the band energy for conduction electrons, $V_{\mathbf{k}}$ is the matrix
element of hybridization. It is suitable to use the representation of the
Fermi operators for the localized electrons in terms of the many-body atomic
quantum numbers \cite{ii}
\begin{equation}
a_{il\sigma m}^{\dagger }=(n+1)^{1/2}\sum G_{SL}^{S^{\prime }L^{\prime
}}C_{L\mu ,lm}^{L^{\prime }\mu ^{\prime }}C_{SM,\frac 12\sigma }^{S^{\prime
}M^{\prime }}X_i\left( S^{\prime }L^{\prime }M^{\prime }\mu ^{\prime
},SLM\mu \right)  \label{pro}
\end{equation}
where $G_{SL}^{S^{\prime }L^{\prime }}$ are the fractional parentage
coefficients, $X_i(\alpha \beta )=|i\alpha \rangle \langle i\beta |$ are the
many-electron Hubbard operators \cite{Hubbard-IV:1965} which satisfy the
relations at a site $i$%
\begin{equation}
X_i(\alpha \beta )X_i(\gamma \varepsilon )=\delta _{\beta \gamma }X_i(\alpha
\varepsilon ),\,\,\sum_\alpha X_i(\alpha \alpha )=1.  \label{kin}
\end{equation}

By a canonical Schrieffer-Wolff transformation the model (\ref{m1}) is
reduced to the broad-band $s-d$ exchange model with the exchange parameter $%
I_{sd}\sim V^2/\Delta $ ($\Delta $ is the difference of the term energies).
In a general case, we have the $s-d$ exchange model with two magnetic
configurations. This model was used to investigate the electron spectrum and
calculate the Kondo temperature in Refs.\onlinecite{ii,ii93}. Here we
introduce a further simplification assuming that only one of the competing
configurations has non-zero orbital moment $L=l$. This assumption holds for
the magnetite which has $d^5$ and $d^6$ ground-state configurations for Fe$%
^{3+}$ and Fe$^{2+}$ respectively, the first configuration having zero
orbital moment. A similar situation takes place for the CMR manganites (with
$d^3$ and $d^4$ configuration for Mn$^{4+}$ and Mn$^{3+}$ ): due to the
relevance of $t_{2g}-e_g$ crystal-field splitting the former configuration
corresponds to the completely filled $t_{2g}$ band with $L=0$.

According to Refs.\onlinecite{ii,ii93} the $s-d$ exchange Hamiltonian takes
the form
\begin{equation}
\mathcal{H}_{sd}=-I_{sd}\sum C_{SM,\frac 12\sigma }^{S^{\prime }M^{\prime
}}C_{SM^{\prime \prime },\frac 12\sigma ^{\prime }}^{S^{\prime }M^{\prime
}}X_i(SLMm,SLM^{\prime \prime }m)c_{i\sigma m}^{\dagger }c_{i\sigma ^{\prime
}m}.  \label{h2}
\end{equation}
Rearranging the product of the Clebsch-Gordan coefficients $C$ with the use
of $6j$-symbols,
\begin{equation}
\sum_{M^{\prime }}C_{SM,\frac 12\sigma }^{S^{\prime }M^{\prime
}}C_{SM^{\prime \prime },\frac 12\sigma ^{\prime }}^{S^{\prime }M^{\prime }}=%
\frac{3\sqrt{2}}2\left( \frac{2S^{\prime }+1}{2S+1}\right) ^{1/2}\left\{
\begin{array}{ccc}
1/2 & 1/2 & 1 \\
S & S^{\prime } & S^{\prime }
\end{array}
\right\} \sum_{q=-1}^1C_{SM^{\prime \prime },1q}^{SM}C_{\frac 12\sigma
^{\prime },1-q}^{\frac 12\sigma },  \label{id}
\end{equation}
and taking into account that $C_{SM^{\prime \prime },1q}^{SM}$ are the
matrix elements of spin operators, we obtain the $s-d$ exchange Hamiltonian
with orbital degrees of freedom in the form
\begin{equation}
\mathcal{H}=\sum_{\mathbf{k}\sigma m}t_{\mathbf{k}m}c_{\mathbf{k}\sigma
m}^{\dagger }c_{\mathbf{k}\sigma m}-I\sum_{i\sigma \sigma ^{\prime }m}%
\mathbf{S}_i\mbox {\boldmath $\sigma $}_{\sigma \sigma ^{\prime }}c_{i\sigma
m}^{\dagger }c_{i\sigma ^{\prime }m}+\mathcal{H}_d  \label{H0}
\end{equation}
where $\mathcal{H}_d$ is the Heisenberg Hamiltonian of the
localized-spin system, {\boldmath $\sigma $} are the Pauli
matrices, the $s-d$ exchange parameter $I$ is renormalized in
comparison with $I_{sd}$ in Eq.(\ref{h2}). A similar one-impurity
model was discussed for the Mn$^{2+}$ ion (orbital singlet
problem) \cite{Hewson,Tsvelick}.

In contrast with the case of rare-earth elements where the orbital moments
are not quenched, but act as a part of the total orbital moment (the
Russel-Saunders coupling \cite{i,ii,ii93}), for 3$d$-electron systems we
have to consider the case of strong crystal field where the excitation
spectrum is characterized by spin and orbital quantum numbers separately.
However, in the case under consideration the quantum number $m$ enters the
model in a very simple way (the Hamiltonian is diagonal in the orbital
indices) and can label both spherical and cubic harmonics. Of course, a
relatively simple form of the Hamiltonian (\ref{H0}) results from our
assumption that only one configuration has a non-zero orbital moment.
Fortunately, it is the case that is most interesting for real systems, as it
was mentioned above. Thus a complicated mathematics involving the orbital
Clebsch-Gordan coefficients is relevant only when we have \textit{two}
many-electron terms with $L\ne 0$.

Note that in the case of $j-j$ coupling (e.g., 5$f$-electron systems) where
the electron states $|jm\rangle $ are characterized by the projection of the
momentum $j$, we have the Hamiltonian
\begin{eqnarray}
\mathcal{H} &=&\sum_{\mathbf{k}m}t_{\mathbf{k}m}c_{\mathbf{k}m}^{\dagger }c_{%
\mathbf{k}m}-I\sum_i\mathbf{J}_i\mathbf{j}_i,  \nonumber \\
j_i^\alpha &=&\sum_{mm^{\prime }}C_{jm,1\alpha }^{jm^{\prime
}}c_{im}^{\dagger }c_{im^{\prime }}.
\end{eqnarray}

Provided that the crystal field Hamiltonian is diagonal in $m$,
the model (\ref{H0}) does not mix (at least in the lowest orders
of perturbation theory) the states with different $m$, see Section
4. Therefore we focus in the present work mainly on a more rich
and complicated narrow-band case (which is also most interesting
for real transition metal compounds). This case should be
described by a two-configuration Hubbard model where both
conduction electrons and local moments belong to the same
$d$-band, the states with $n+1$ electrons playing the role of
current-carrier states. After performing the procedure of mapping
onto the corresponding state space, the one-electron Fermi
operators for the strongly correlated states $a_{il\sigma
m}^{\dagger }$ are replaced by many-electron operators according
to Eq.(\ref{pro}). Taking into account the values of the
Clebsh-Gordan coefficients which correspond to the coupling of
momenta $S$ and 1/2 we obtain
\begin{equation}
\mathcal{H}=\sum_{\mathbf{k}\sigma m}t_{\mathbf{k}m}g_{\mathbf{k}\sigma
m}^{\dagger }g_{\mathbf{k}\sigma m}.  \label{eq:H}
\end{equation}
Here we have redefined the band energy by including the many-electron
renormalization factor,
\begin{equation}
t_{\mathbf{k}m}(n+1)(G_{SL}^{S^{\prime }0})^2/(2l+1)\rightarrow t_{\mathbf{k}%
m},  \label{ren}
\end{equation}
and
\begin{eqnarray}
g_{i\sigma m}^{\dagger } &=&\sum_{M=-S}^S\sqrt{\frac{S-\sigma M}{2S+1}}%
X_i(S-1/2,M+\frac \sigma 2;SMm),\;\;\;S^{\prime }=S-1/2,  \nonumber \\
g_{i\sigma m}^{\dagger } &=&\sum_{M=-S}^S\sqrt{\frac{S+\sigma M+1}{2S+1}}%
X_i(S+1/2,M+\frac \sigma 2;SMm),\;\;\;S^{\prime }=S+1/2
\end{eqnarray}
where $|SMm\rangle $ are the empty states with the orbital index $m,$ $%
|S^{\prime }M^{\prime }\rangle $ are the singly-occupied states with the
total on-site spin $S^{\prime }=S\pm 1/2$ and its projection $M^{\prime }$, $%
\sigma =\pm $.

We see that the two-configuration Hamiltonian is a generalization of the
narrow-band $s-d$ exchange model with $|I|\rightarrow \infty $ \cite
{ii,jpc85}. In the case where the configuration $d^{n+1}$ has larger spin
than the configuration $d^n$, we have the effective ``$s-d$ exchange model''
with $I>0$, and in the opposite case with $I<0$; it is worthwhile to remind
that our band energy is renormalized according to (\ref{ren}) in comparison
with the $s-d$ exchange Hamiltonian (cf. the factor of 1/2 for the
narrow-band Hubbard model without orbital degeneracy \cite{jpc85}).

In the case of $j-j$ coupling we have
\[
g_{i\sigma m}^{\dagger }\rightarrow g_{im}^{\dagger }=\sum_{MM^{\prime
}}C_{JM,jm}^{J^{\prime }M^{\prime }}X_i(J^{\prime }M^{\prime },JM)
\]

The specification of the Hamiltonian (\ref{eq:H}) for the case of a
saturated ferromagnetic state (where only maximum spin projections give a
contribution) reads for $I<0$ ($S^{\prime }=S-1/2$)
\begin{equation}
\mathcal{H}=\sum_{\mathbf{k}m}t_{\mathbf{k}m}\left[ \frac{2S}{2S+1}X_{-%
\mathbf{k}}^{a,+m}X_{\mathbf{k}}^{+m,a}+\frac 1{2S+1}X_{-\mathbf{k}%
}^{a,-m}X_{\mathbf{k}}^{-m,a}+\frac{\sqrt{2S(2S-1)}}{2S+1}\left( X_{-\mathbf{%
k}}^{a,+m}X_{\mathbf{k}}^{-m,b}+\mathrm{h.c.}\right) \right] .  \label{exp1}
\end{equation}
Here $|+m\rangle $ and $|-m\rangle $ are the on-site states without excess
electrons and with localized spin projection $S$ and $S-1$, respectively; $%
|a\rangle =$ $|S-1/2,S-1/2\rangle $ and $|b\rangle =$
$|S-1/2,S-3/2\rangle $ are the states with one excess electron and
the total spin projection $S-1/2$ and $S-3/2$, respectively;
$X_{\mathbf{k}}^{\alpha \beta }$ are the Fourier transforms of the
Hubbard operators $X_{i}(\alpha, \beta)$.

For the case where $I>0$ ($S^{\prime }=S+1/2$) we have
\begin{equation}
\mathcal{H}=\sum_{\mathbf{k}m}t_{\mathbf{k}m}\left[ X_{-\mathbf{k}}^{u,+m}X_{%
\mathbf{k}}^{+m,u}+\frac{2S}{2S+1}X_{-\mathbf{k}}^{v,-m}X_{\mathbf{k}%
}^{-m,v}+\frac 1{2S+1}X_{-\mathbf{k}}^{v,+m}X_{\mathbf{k}}^{+m,v}+\sqrt{%
\frac{2S}{2S+1}}\left( X_{-\mathbf{k}}^{u,+m}X_{\mathbf{k}}^{-m,v}+\mathrm{%
h.c.}\right) \right]   \label{exp2}
\end{equation}
where $|u\rangle =$ $|S+1/2,S+1/2\rangle $ and $|v\rangle =$ $%
|S+1/2,S-1/2\rangle $ are the singly-occupied states.

We will consider also the simplest model where one of the configuration
corresponding to the ``current carriers'' has zero spin and orbital moments
(non-magnetic holes):

\begin{equation}
\mathcal{H}=\sum_{\mathbf{k}\sigma m}t_{\mathbf{k}m}X_{-\mathbf{k}%
}^{0,\sigma m}X_{\mathbf{k}}^{\sigma m,0}.  \label{HHM}
\end{equation}
This is a formal generalization of the standard narrow-band
Hubbard model in the $X$-operator representation to include the
orbital degeneracy. Apart from the $m$-dependence of the band
energy $t_{\mathbf{k}m}$, the Hamiltonian (\ref{HHM}) includes
spin and orbital degrees of freedom on equal footings. However,
for finite Hubbard repulsion $U$ the symmetry of spin and charge
channels is in fact lost owing to the Hund interaction, so that
the state with orbital ferromagnetism and spin antiferromagnetism
can become favorable in comparison with the spin antiferromagnetic
state \cite{kugel}. Note the equivalence of the model (\ref{HHM})
with the replacement $t_{\mathbf{k}m}\rightarrow
t_{\mathbf{k}m}/2$ and the model (\ref{exp1}) for $S=1/2$.

For orbitally degenerate case the subbands $t_{\mathbf{k}m}$ are connected
by the point-group symmetry transformations and have identical densities of
states. However, in real situations usually the Jahn-Teller lattice
instability takes place which destroys the exact orbital degeneracy \cite
{kugel}. Due to the Jahn-Teller distortions, both additional on-site
crystal-field splitting arises (which means the shift of the centers of the
subbands) and hopping integrals are changed (which means different $\mathbf{k%
}$-dependences). Thus we will assume further that all the functions $t_{%
\mathbf{k}m}$ are in general different.

\section{Collective spin and orbital excitations}

The spectrum of electron and spin excitations in the above-discussed narrow
band models can be calculated similar to the papers \cite{ai88,jpc85,iz04},
the difference being in the occurrence of transitions into orbital states.

Even in the narrow band case where small interaction parameter is absent we
have an hierarchy of energy scales owing to both formal expansion parameters
(inverse nearest-neighbor number $1/z$, quasiclassical parameter $1/S$) and
small current-carrier concentration. Therefore the width of the ``magnon''
band is much smaller than characteristic electron energies. A similar
approach was developed earlier for degenerate ferromagnetic semiconductors
\cite{jpc85}.

First we calculate the retarded commutator Green's function describing spin
and orbital excitations in the simplest model (\ref{HHM}),
\[
G_{\mathbf{q}mm^{\prime }}^{\sigma \sigma ^{\prime }}(\omega )=\langle
\langle X_{\mathbf{q}}^{\sigma m,\sigma ^{\prime }m^{\prime }}|X_{-\mathbf{q}%
}^{\sigma ^{\prime }m^{\prime },\sigma m}\rangle \rangle _\omega ,\,\mathrm{%
Im}\omega >0.
\]
To derive the equation of motion, we use the commutation rules for
$X$ -operators, which follow from Eq.(\ref{kin}). We obtain
\begin{equation}
\omega G_{\mathbf{q}mm^{\prime }}^{\sigma \sigma ^{\prime
}}(\omega
)=N_{\sigma m}-N_{\sigma ^{\prime }m^{\prime }}+\sum_{\mathbf{k}}(t_{\mathbf{%
k-q}m^{\prime }}-t_{\mathbf{k}m})\langle \langle X_{\mathbf{q}-\mathbf{k}%
}^{0,\sigma ^{\prime }m^{\prime }}X_{\mathbf{k}}^{\sigma m,0}|X_{-\mathbf{q}%
}^{\sigma ^{\prime }m^{\prime },\sigma m}\rangle \rangle _{\omega
}. \label{e1}
\end{equation}
In the next equation of motion we make a simple decoupling described in Ref.%
\onlinecite{ai88}, which corresponds to the first-order term in $1/z$%
-perturbation theory (each order in $1/z$ corresponds formally to
a summation over an additional wave vector):
\begin{eqnarray}
(\omega -E_{\mathbf{k}\sigma m}+E_{\mathbf{k}-\mathbf{q}\sigma
^{\prime }m^{\prime }})\langle \langle
X_{\mathbf{q}-\mathbf{k}}^{0,\sigma ^{\prime }m^{\prime
}}X_{\mathbf{k}}^{\sigma m,0}|X_{-\mathbf{q}}^{\sigma ^{\prime
}m^{\prime },\sigma m}\rangle \rangle _{\omega }
&=&n_{\mathbf{k}\sigma
m}-n_{\mathbf{k-q}\sigma ^{\prime }m^{\prime }}  \nonumber \\
&&\ \ \ \ \ +(t_{\mathbf{k}m}n_{\mathbf{k}\sigma
m}-t_{\mathbf{k-q}m^{\prime }}n_{\mathbf{k-q}\sigma ^{\prime
}m^{\prime }})G_{\mathbf{q}mm^{\prime }}^{\sigma \sigma ^{\prime
}}(\omega ).  \label{e2}
\end{eqnarray}
Here the ``Hubbard-I'' energies $E_{\mathbf{k}\sigma m}$ and the occupation
numbers $n_{\mathbf{k}\sigma m}$ are given by the ``Hubbard-I'' \cite{hubI}
expressions
\begin{eqnarray}
E_{\mathbf{k}\sigma m} &=&t_{\mathbf{k}m}(N_{\sigma m}+N_0),\,\,N_\alpha
=\langle X_i^{\alpha \alpha }\rangle ,  \nonumber \\
n_{\mathbf{k}\sigma m} &=&\langle X_{-\mathbf{k}}^{0,\sigma m}X_{\mathbf{k}%
}^{\sigma m,0}\rangle =(N_{\sigma m}+N_0)f(E_{\mathbf{k}\sigma m}),
\end{eqnarray}
$f(E)$ being the Fermi function. Then we have
\begin{eqnarray}
G_{\mathbf{q}mm^{\prime }}^{\sigma \sigma ^{\prime }}(\omega ) &=&\left(
N_{\sigma m}-N_{\sigma ^{\prime }m^{\prime }}+\sum_{\mathbf{k}}\frac{(t_{%
\mathbf{k-q}m^{\prime }}-t_{\mathbf{k}m})(n_{\mathbf{k}\sigma m}-n_{\mathbf{%
k-q}\sigma ^{\prime }m^{\prime }})}{\omega -E_{\mathbf{k}\sigma m}+E_{%
\mathbf{k}-\mathbf{q}\sigma ^{\prime }m^{\prime }}}\right)  \nonumber \\
&&\ \ \ \times \left( \omega -\sum_{\mathbf{k}}\frac{(t_{\mathbf{k-q}%
m^{\prime }}-t_{\mathbf{k}m})(t_{\mathbf{k}m}n_{\mathbf{k}\sigma m}-t_{%
\mathbf{k-q}m^{\prime }}n_{\mathbf{k-q}\sigma ^{\prime }m^{\prime }})}{%
\omega -E_{\mathbf{k}\sigma m}+E_{\mathbf{k}-\mathbf{q}\sigma ^{\prime
}m^{\prime }}}\right) ^{-1}.  \label{gh}
\end{eqnarray}
In the saturated ferromagnetic state ($N_{\sigma m}=\delta _{\sigma +}N_m$%
) and for small hole concentrations $c=N_0$ the pole of the Green's function
(\ref{gh}) yields a simple result for the spin-wave frequency
\begin{equation}
\omega _{\mathbf{q}mm^{\prime }}^{+-}=\sum_{\mathbf{k}}(t_{\mathbf{k}-%
\mathbf{q}m^{\prime }}-t_{\mathbf{k}m})f(t_{\mathbf{k}m}).  \label{om1}
\end{equation}
In the case of saturated orbital ordering ($N_{\sigma m}=\delta
_{m+}N_\sigma $) the orbital excitation frequency $\omega _{\mathbf{q}%
+m^{\prime }}^{\sigma \sigma ^{\prime }}$ ($m=+\neq m^{\prime }$)
is given by the same expression (\ref{om1}). Provided that the
states $m^{\prime }$ and $m$ are split by the crystal field, there
is a gap in the corresponding orbiton spectrum branch. Assuming
for simplicity that the band bottom corresponds to $\Gamma$-point
$\mathbf{k}=0,$ we derive
\begin{equation}
\omega _{\mathbf{q}+m^{\prime }}^{\sigma \sigma ^{\prime }}\simeq c(t_{%
\mathbf{q}m^{\prime }}-t_{0m^{\prime }})+c\Delta ,\,\Delta =t_{m^{\prime
}}^{\min }-t_{+}^{\min }.  \label{om2}
\end{equation}
It should be noted that the gap $\omega _0=c\Delta $ is
proportional to the current carrier concentration, but not just
equal to the crystal-field splitting $\Delta $, in contrast with
the case of weak crystal field typical for the rare-earth systems
\cite{fulde}. Actually, in our case characteristic orbiton
energies for finite wave vectors have the same order of magnitude
as the magnon energies.

The magnon damping can be obtained similar to
Ref.\onlinecite{jpc85} by calculating higher-order terms in $1/z$,
\begin{eqnarray}
\gamma _{\mathbf{q}mm^{\prime }}(\omega ) &=&\pi
\sum_{\mathbf{kp}m^{\prime
\prime }}(t_{\mathbf{k-q}m^{\prime }}-t_{\mathbf{k}m})^{2}[n_{\mathbf{k-q+p}%
m^{\prime \prime }}(1-n_{\mathbf{k}m})+N_{B}(\omega _{\mathbf{p}})(n_{%
\mathbf{k-q+p}m^{\prime \prime }}-n_{\mathbf{k}m})]\delta (\omega +t_{%
\mathbf{k}m}-t_{\mathbf{k-q+p}m^{\prime \prime }}-\omega
_{\mathbf{p}})
\nonumber \\
&=&\pi \sum_{\mathbf{kp}m^{\prime \prime }}(t_{\mathbf{k-q}m^{\prime }}-t_{%
\mathbf{k}m})^{2}(\omega _{\mathbf{p}}-\omega )\frac{\partial n_{\mathbf{k}m}%
}{\partial t_{\mathbf{k}m}}[N_{B}(\omega _{\mathbf{p}})-N_{B}(\omega _{%
\mathbf{p}}-\omega )]\delta
(t_{\mathbf{k}m}-t_{\mathbf{k-q+p}m^{\prime \prime }})
\end{eqnarray}
with $N_{B}(\omega _{\mathbf{p}})$ being the Bose function. The
damping turns out to be finite at $T=0$, unlike the case of the
Heisenberg ferromagnet. However, it is proportional to the
frequency $\omega $ and
contains formal small parameters, which leads to very small value \cite%
{jpc85}. Thus both spin-flip and non-spin-flip orbital excitations
are well defined. This fact is non-trivial in a narrow-band case
(a similar situation takes place also in the antiferromagnetic
case, cf. Ref.\onlinecite{kikoin}).

In the model (\ref{eq:H}) the results for the ``magnon'' spectrum are more
complicated and depend in a non-trivial way on the spin value. The
calculations for the ferromagnetically saturated state can be performed
similar to Ref.\onlinecite{jpc85}. For the case $S^{\prime }=S-1/2$ we write
down the sequence of equations of motion
\begin{eqnarray}
\omega G_{\mathbf{q}mm^{\prime }}^{+-}(\omega ) &=&N_m+\frac 1{2S+1}\sum_{%
\mathbf{k}}[(t_{\mathbf{k}-\mathbf{q}m^{\prime }}-2St_{\mathbf{k}m})\langle
\langle X_{\mathbf{q}-\mathbf{k}}^{a,-m^{\prime }}X_{\mathbf{k}}^{+m,a}|X_{-%
\mathbf{q}}^{-m^{\prime },+m}\rangle \rangle _\omega  \nonumber \\
&&\ \ \ \ \ \ \ \ \ \ \ \ \ \ +\sqrt{2S(2S-1)}t_{\mathbf{k}m^{\prime
}}\langle \langle X_{-\mathbf{k}}^{a,+m^{\prime }}X_{\mathbf{k+q}%
}^{+m,b}|X_{-\mathbf{q}}^{-m^{\prime },+m}\rangle \rangle _\omega ].
\end{eqnarray}
The equation for the first Green's function in the right-hand side is
obtained as above (see Eq.(\ref{e2})), and for the second one we have
\begin{equation}
(\omega -t_{\mathbf{k}m^{\prime }}^{*}N_m)\langle \langle X_{-\mathbf{k}%
}^{a,+m^{\prime }}X_{\mathbf{k+q}}^{+m,b}|X_{-\mathbf{q}}^{-m^{\prime
},+m}\rangle \rangle _\omega =\sqrt{2S(2S-1)}\langle X_{-\mathbf{k}%
}^{a,+m}X_{\mathbf{k}}^{+m,a}\rangle t_{\mathbf{k}m^{\prime }}G_{\mathbf{q}%
mm^{\prime }}^{+-}(\omega )
\end{equation}
where
\begin{equation}
t_{\mathbf{k}m}^{*}=\frac{2S}{2S+1}t_{\mathbf{k}m}.  \label{t*}
\end{equation}
Then the excitation energy in the leading order in $1/z$ reads
\begin{equation}
\omega _{\mathbf{q}mm^{\prime }}^{+-}=\frac 1{2S}\sum_{\mathbf{k}}[(t_{%
\mathbf{k}-\mathbf{q}m^{\prime }}^{*}-2St_{\mathbf{k}m}^{*})f(t_{\mathbf{k}%
m}^{*})+(2S-1)t_{\mathbf{k}m^{\prime }}^{*}f(t_{\mathbf{k}m^{\prime }}^{*})]
\label{ss}
\end{equation}
(more strict calculations in the absence of orbital degeneracy \cite{jpc85}
yield a closed integral equation for the magnon spectrum). At the same time,
the frequency of the non-spin-flip orbital transitions is obtained similar
to Eq.(\ref{om1}) and is given for a saturated orbital-ordered state by the
expression
\begin{equation}
\omega _{\mathbf{q}mm^{\prime }}^{++}=\sum_{\mathbf{k}}(t_{\mathbf{k}-%
\mathbf{q}m^{\prime }}^{*}-t_{\mathbf{k}m}^{*})f(t_{\mathbf{k}m}^{*}),
\end{equation}
so that a dependence on $S$ is absent.

For $S^{\prime }=S+1/2$ we have the equations of motion
\begin{equation}
\omega G_{\mathbf{q}mm^{\prime }}^{+-}(\omega )=N_m-\frac 1{2S+1}\sum_{%
\mathbf{k}}t_{\mathbf{k}m}\langle \langle X_{\mathbf{q}-\mathbf{k}%
}^{u,-m^{\prime }}X_{\mathbf{k}}^{+m,u}-\sqrt{\frac{2S}{2S+1}}X_{-\mathbf{k}%
}^{u,+m}X_{\mathbf{k+q}}^{+m^{\prime },v}|X_{-\mathbf{q}}^{-m^{\prime
},+m}\rangle \rangle _\omega ,
\end{equation}
\begin{equation}
(\omega -t_{\mathbf{k}m}N_m)\langle \langle X_{\mathbf{q}-\mathbf{k}%
}^{u,-m^{\prime }}X_{\mathbf{k}}^{+m,u}|X_{-\mathbf{q}}^{-m^{\prime
},+m}\rangle \rangle _\omega =-t_{\mathbf{k}m}\langle X_{-\mathbf{k}%
}^{u,+m}X_{\mathbf{k}}^{+m,u}\rangle G_{\mathbf{q}mm^{\prime }}^{+-}(\omega
),
\end{equation}
\begin{equation}
\lbrack \omega +t_{\mathbf{k}m}N_m-t_{\mathbf{k}+\mathbf{q}m^{\prime
}}N_{m^{\prime }}/(2S+1)]\langle \langle X_{-\mathbf{k}}^{u,+m}X_{\mathbf{k+q%
}}^{+m^{\prime },v}|X_{-\mathbf{q}}^{-m^{\prime },+m}\rangle \rangle _\omega
=\sqrt{\frac{2S}{2S+1}}\langle X_{-\mathbf{k}}^{u,+m}X_{\mathbf{k}%
}^{+m,u}\rangle G_{\mathbf{q}mm^{\prime }}^{+-}(\omega ).
\end{equation}
Then we derive
\begin{equation}
\omega _{\mathbf{q}mm^{\prime }}^{+-}=\sum_{\mathbf{k}}\frac{t_{\mathbf{k}+%
\mathbf{q}m^{\prime }}N_{m^{\prime }}-t_{\mathbf{k}m}N_m}{2St_{\mathbf{k}%
m}N_m+t_{\mathbf{k}m}N_m-t_{\mathbf{k}+\mathbf{q}m^{\prime }}N_{m^{\prime }}}%
t_{\mathbf{k}m}f(t_{\mathbf{k}m}).  \label{w3}
\end{equation}
The orbital frequency for a saturated state reads
\begin{equation}
\omega _{\mathbf{q}mm^{\prime }}^{++}=\sum_{\mathbf{k}}(t_{\mathbf{k+q}%
m^{\prime }}-t_{\mathbf{k}m})f(t_{\mathbf{k}m}).
\end{equation}

It should be mentioned that, instead of the difference $t_{\mathbf{k}%
m}N_m-t_{\mathbf{k}+\mathbf{q}m^{\prime }}N_{m^{\prime }},$ a resolvent
occurs in the denominator in the right-hand side of Eq.(\ref{w3}) at more
rigorous calculations which do not use the $1/z$-expansion (cf. Ref.%
\onlinecite{jpc85}). In particular, in the case of
ferromagnetically saturated spin and orbital ordering (all the
sites without excess electrons are in the orbital state
$|+m\rangle =|++\rangle $ with some $m\equiv +$) we obtain the
following exact result for the magnon pole ($m=m^{\prime }$):
\begin{eqnarray}
\omega _{\mathbf{q}m}^{+-} &=&\sum_{\mathbf{k}}\frac{t_{\mathbf{k}+\mathbf{q}%
m}-t_{\mathbf{k}m}}{2S+(t_{\mathbf{k}m}-t_{\mathbf{k}+\mathbf{q}m})R_{%
\mathbf{k}m}(\omega _{\mathbf{q}m}^{+-})}f(t_{\mathbf{k}m}),  \label{exact}
\\
R_{\mathbf{k}m}(\omega ) &=&\sum_{\mathbf{p}}\frac{1-f(t_{\mathbf{k+p}m})}{%
\omega +t_{\mathbf{k}m}-t_{\mathbf{k}+\mathbf{p}m}-\omega _{\mathbf{p}m}^{+-}%
}.
\end{eqnarray}
Note that the imaginary part of the expression (\ref{exact}) describes the
magnon damping.

\section{Electron spectrum and instabilities of the saturated state}

In the broad-band $s-d$ exchange model (\ref{H0}) we can perform a
decoupling in spirit of a ladder approximation (cf. Ref.\onlinecite{ai}) to
obtain the Green's function for the case of zero temperature $T=0$%
\begin{equation}
\langle \!\langle c_{\mathbf{k}\sigma m}|c_{\mathbf{k}\sigma m}^{\dagger
}\rangle \!\rangle _E=\left[ E-t_{\mathbf{k}\sigma m}-\Sigma _{\mathbf{k}%
m}^\sigma (E)\right] ^{-1}  \label{g0}
\end{equation}
where $t_{\mathbf{k}\sigma m}=t_{\mathbf{k}m}-\sigma IS$ is the mean-field
(Hartree-Fock) spectrum. The electron self-energy reads
\begin{eqnarray}
\Sigma _{\mathbf{k}m}^\sigma (E) &=&\frac{2I^2SR_{\mathbf{k}m}^\sigma }{%
1+\sigma IR_{\mathbf{k}m}^\sigma },  \label{g1} \\
R_{\mathbf{k}m}^{\uparrow }(E) &=&\sum_{\mathbf{q}}\frac{n_{\mathbf{k-q}%
m}^{\downarrow }}{E-t_{\mathbf{k-q\downarrow }m}+\omega _{\mathbf{q}}},R_{%
\mathbf{k}}^{\downarrow }(E)=\sum_{\mathbf{q}}\frac{1-n_{\mathbf{k-q}%
m}^{\uparrow }}{E-t_{\mathbf{k-q\uparrow }m}-\omega _{\mathbf{q}}}
\label{g2}
\end{eqnarray}
with $n_{\mathbf{k}m}^\sigma =f(t_{\mathbf{k}\sigma m})$.
Imaginary part of (\ref{g1}) describes the non-quasiparticle
states which are due to non-pole contributions (branch cut of the
Green's function) \cite{UFN,ikjpc90}. In the saturated
ferromagnetic case $n_{\mathbf{k}m}^{\downarrow }\neq 0$ for $I<0$
and $n_{\mathbf{k}m}^{\uparrow }\neq 0$ for $I>0,$ and these
states with $\sigma =-\mathrm{sign}I$ lie below (above) the Fermi
level, respectively. The corresponding densities of states for a
twofold orbital-degenerate band are shown in Figs.1,2. One can see
that in the model (\ref{H0}) the subbands with different $m$ are
uncoupled (we do not consider their mixing owing to off-diagonal
crystal field), so that the picture of the half-metallic
ferromagnetism state discussed in Refs.\onlinecite{UFN,ikjpc90} is
not qualitatively changed, although the number of excitation
branches increases.

A more complicated situation takes place for the narrow-band limit, a formal
reason being in occurrence of inter-subband transitions owing to non-trivial
commutation relations for many-electron $X$-operators. First we consider the
one-particle spectrum in the simplest model (\ref{HHM}). We treat the case $%
T=0$ and restrict ourselves to the case of spin- and orbital-saturated
ferromagnetic state (all the singly occupied on-site states have the spin
projection $\sigma =+$ and the orbital state $m\equiv +$). Since this ground
state is non-degenerate, the one-particle spectrum at small hole
concentrations $c=N_0$ can be investigated in a strict way similar to Ref.%
\onlinecite{jpc85}. We have to calculate the retarded anticommutator Green's
function
\begin{equation}
G_{\mathbf{k}\sigma m}(E)=\langle \!\langle X_{\mathbf{k}}^{\sigma m,0}|X_{-%
\mathbf{k}}^{0,\sigma m}\rangle \!\rangle _E,\qquad \mathrm{Im}E>0.
\label{eq:EGF}
\end{equation}
The current carriers with $|\sigma m\rangle =|++\rangle $ propagate as free
ones,
\begin{equation}
G_{\mathbf{k}++}(E)=(E-t_{\mathbf{k}+})^{-1}.
\end{equation}
In the case $|\sigma m\rangle \neq |++\rangle ,$ to extract the
operators of spin and orbital excitations, it is convenient to use
the kinematical relation which follows from Eq.(\ref{kin})
\begin{equation}
X_{\mathbf{k}}^{\sigma m,0}=\sum_{\mathbf{q}}X_{\mathbf{q}}^{\sigma m,++}X_{%
\mathbf{k-q}}^{++,0}.
\end{equation}
Then we have
\begin{equation}
G_{\mathbf{k}\sigma m}(E)=\sum_{\mathbf{q}}F_{\mathbf{kq}\sigma
m}(E),\,\;\;\;F_{\mathbf{kq}\sigma m}(E)=\langle \!\langle X_{\mathbf{q}%
}^{\sigma m,++}X_{\mathbf{k-q}}^{++,0}|X_{-\mathbf{k}}^{0,\sigma m}\rangle
\!\rangle _E.
\end{equation}
After a natural decoupling procedure (cf. Ref.\onlinecite{ikjpc90}) we
obtain the integral equation for the function $F$
\begin{equation}
(E-t_{\mathbf{k-q}+}+\omega _{\mathbf{q}+m}^{+\sigma })F_{\mathbf{kq}\sigma
m}(E)=n_{\mathbf{k}-\mathbf{q}}[1-(t_{\mathbf{k-q}+}-t_{\mathbf{k}m})\sum_{%
\mathbf{p}}F_{\mathbf{kp}\sigma m}(E)]
\end{equation}
where $n_{\mathbf{k}}=$ $\langle X_{-\mathbf{k}}^{0,++}X_{\mathbf{k}%
}^{++,0}\rangle \!=f(t_{\mathbf{k}+}).$ Solving this equation we derive
\begin{equation}
G_{\mathbf{k}\sigma m}(E)=\frac 1{E-t_{\mathbf{k}m}+2\overline{S}/R_{\mathbf{%
k}\sigma m}(E)}  \label{g11}
\end{equation}
where $2\overline{S}=1-c$ is the average spin magnetization,
\begin{equation}
R_{\mathbf{k}\sigma m}(E)=\sum_{\mathbf{q}}\frac{n_{\mathbf{k}-\mathbf{q}}}{%
E-t_{\mathbf{k-q}+}+\omega _{\mathbf{q}+m}^{+\sigma }}.  \label{l14}
\end{equation}
Thus the Green's function with minority spin (orbital) projection has
non-pole structure, the non-quasiparticle states below the Fermi level being
of crucial importance to satisfy the sum rule
\begin{eqnarray}
\int dEf(E)g_{\sigma m}(E) &=&\sum_{\mathbf{k}}\langle X_{-\mathbf{k}%
}^{0,\sigma m}X_{\mathbf{k}}^{\sigma m,0}\rangle =N_0,  \nonumber \\
g_{\sigma m}(E) &=&-\frac 1\pi \mathrm{Im}\sum_{\mathbf{k}}G_{\mathbf{k}%
\sigma m}(E),
\end{eqnarray}
irrespective of $\sigma ,$ $m.$

The spectrum picture is presented in Fig.3. To simplify numerical
calculations, we average the resolvent $R_{\mathbf{k}\sigma m}(E)$
in (\ref {g11}) over $\mathbf{k,}$
\[
R_{\mathbf{k}\sigma m}(E)\rightarrow \overline{R}_{\sigma m}(E)=\int d\omega
K_{\sigma m}(\omega )\sum_{\mathbf{k}}\frac{n_{\mathbf{k}}}{E-t_{\mathbf{k}%
+}+\omega }
\]
(this approximation yields the correct behavior near the Fermi
level, cf. Ref.\onlinecite{iz04}, although yields an irrelevant
unphysical shift of the band bottom by the maximum magnon
frequency). We use the semielliptic magnon density of states
$K_{\sigma m}(\omega )$ which is proportional (with the
corresponding shift) to the bare electron density of states
according to (\ref{om2}), the band edge $\omega _0$ being
determined by the crystal-filed splitting. When neglecting spin
(orbital) dynamics, $\mathrm{Im}R_{\mathbf{k}\sigma m}(E)$
has a jump at the Fermi level. For gapless magnons $\mathrm{Im}\overline{R}%
_{\sigma m}(E)$ (this quantity is also shown in Fig.3) vanishes at $E_F$
according to the law $(E_F-E)_{}^{3/2}$ \cite{ikjpc90}. The ``Kondo''
singularity of $\mathrm{Re}\overline{R}_{\sigma m}(E)$ owing to the Fermi
function $n_{\mathbf{k}}$ results in that the dependence of $g_{\sigma m}(E)$
near $E_F$ is considerably more sharp than just of $\mathrm{Im}\overline{R}%
_{\sigma m}(E)$. The contributions of orbitons with a gap spectrum
demonstrate  a threshold energy equal to the gap, that is, in the
presence of the orbital splitting $\mathrm{Im}\overline{R}_{\sigma
m}(E)$ starts from $E_F-\omega _0.$ At the same time, the
threshold energy decreases strongly for the total density of
states $g_{\sigma m}(E)$. Besides that, the orbital splitting
results in a height increase of the $g_{\sigma m}(E)$ peak below
the Fermi level.

The instability of the saturated state ($N_{\sigma m}\neq 0$ for $|\sigma
m\rangle \neq |++\rangle $) with increasing $c$ corresponds to occurrence of
a non-zero spectral density above $E_F$ since
\begin{equation}
\int dE[1-f(E)]g_{\sigma m}(E)=N_{\sigma m}.
\end{equation}
This spectral density comes from the quasiparticle pole determined by the
equation
\begin{equation}
R_{\mathbf{k}\sigma m}(E_F)=\frac{2\overline{S}}{\max t_{\mathbf{k}m}-E_F}.
\end{equation}
The analytical estimations are complicated due to the logarithmic
singularity in the resolvent $R_{\mathbf{k}\sigma m}(E).$ It should be noted
that a simple cutoff of this singularity at a characteristic magnon
frequency (cf. Ref. \cite{edw}) is insufficient since the energy dependence
of the resolvent plays an important role ($\mathrm{Im}R_{\mathbf{k}\sigma
m}(E)\propto (E_F-E)_{}^{3/2}$ and the maximum in $\mathrm{Re}R_{\mathbf{k}%
\sigma m}(E)$ does not correspond to $E_F$). Thus an integration with the
magnon spectral density is required. As it was demonstrated by the
corresponding numerical calculations for the Hubbard model \cite{iz04}, the
critical concentration of holes $c_{\mathrm{crit}}$ makes up about 20-30\%
for a number of crystal lattices. To calculate the second critical
concentrations $c_{\mathrm{crit}}^{\prime }$ (a transition from the
non-saturated ferromagnetism into the paramagnetic state), more advanced
approximations taking into account longitudinal spin fluctuations are
necessary \cite{iz04}. Note that in a more realistic case of finite on-site
interaction the conditions for a ferromagnetically saturated state can be
considerably more strict.

Provided that the spectra $t_{\mathbf{k}m}$ with different $m$ are
connected by crystal symmetry transformations (e.g., states
corresponding to the irreducible $t_{2g}$ representation in a
cubic field), orbital and spin instabilities of the saturated
state coincide. However, if the orbital subbands are split by the
crystal field $\Delta $ (or, in other words, an orbital
pseudomagnetic field is present), the criteria for the orbital and
spin-flip instabilities can become different owing to the
difference in the
spectra $t_{\mathbf{k}m}$ and $t_{\mathbf{k}+}$ and the gap in the spectrum $%
\omega _{\mathbf{q}+m}^{+\sigma }$.

The analytical calculations of the one-particle excitation spectrum for the
model (\ref{exp1}) are performed in a similar way. We restrict ourselves
again to the case of a small conduction electron concentration $c.$ In the
case $S^{\prime }=S-1/2$ spin-down electrons have the spectrum $t_{\mathbf{k}%
+}^{*}$ (see Eq.(\ref{t*})). For the function
\begin{equation}
F_{\mathbf{kq}}(E)=\langle \!\langle X_{\mathbf{q}}^{-,++}X_{\mathbf{k-q}%
}^{++,a}|X_{-\mathbf{k}}^{a,-+}\rangle \!\rangle _E
\end{equation}
we obtain after some simplifications the integral equation
\begin{equation}
(E-t_{\mathbf{k-q}+}^{*}+\omega _{\mathbf{q}++}^{+-})F_{\mathbf{kq}}(E)=n_{%
\mathbf{k}-\mathbf{q}}^{*}[1-\frac 1{2S}(t_{\mathbf{k-q}+}^{*}-t_{\mathbf{k}%
+}^{*})\sum_{\mathbf{p}}F_{\mathbf{kp}}(E)]
\end{equation}
with $n_{\mathbf{k}}^{*}=f(t_{\mathbf{k+}}^{*}).$ Then we derive
\begin{equation}
G_{\mathbf{k}\uparrow +}(E)=\langle \!\langle X_{\mathbf{k}}^{-+,a}|X_{-%
\mathbf{k}}^{a,-+}\rangle \!\rangle _E=\frac{2S}{E-t_{\mathbf{k}%
+}^{*}+(2S-c)/R_{\mathbf{k}\sigma +}^{*}(E)}  \label{l15}
\end{equation}
where $R_{\mathbf{k}\sigma m}^{*}(E)$ differs from (\ref{l14}) by
the replacement $t_{\mathbf{k}+}\rightarrow t_{\mathbf{k}+}^{*}$.
The condition for the spin-flip instability corresponds to
occurrence of the pole of the expression (\ref {l15}) above the
Fermi level. As follows from numerical estimations, such a
condition can be hardly satisfied for $S>1/2$ (although the magnon
frequency is also more soft in this case because of the factor
$(2S)^{-1}$ in Eq.(\ref {ss})).

The orbital instability is essentially governed by the Hamiltonian
of the structure (\ref{HHM}) (the first term in Eq.(\ref{exp1}))
and is determined by the Green's function
\begin{equation}
\langle \!\langle X_{\mathbf{k}}^{+m,a}|X_{-\mathbf{k}}^{a,+m}\rangle
\!\rangle _E=\frac 1{E-t_{\mathbf{k}m}^{*}+(1-c)/R_{\mathbf{k}+m}^{*}(E)}.
\end{equation}
Since in such a case the factor of $2S$ at $1/R^{*}(E)$ is absent, orbital
ordering becomes unstable at some electron concentrations, and we can treat
the state with saturated ferromagnetic spin ordering, but destroyed orbital
ordering.

In the case $S^{\prime }=S+1/2$ (model (\ref{exp2})) spin-up electrons
behave as free ones, and we have for the spin-down Green's function
\begin{eqnarray}
\lbrack E-t_{\mathbf{k}+}^{}/(2S+1)]\langle \!\langle X_{\mathbf{k}%
}^{++,v}|X_{-\mathbf{k}}^{v,++}\rangle \!\rangle _E &=&1-c+\sqrt{\frac{2S}{%
2S+1}}\sum_{\mathbf{q}}t_{\mathbf{k-q+}}L_{\mathbf{kq}}(E), \\
L_{\mathbf{kq}}(E) &=&\langle \langle X_{\mathbf{q}}^{++,-+}X_{\mathbf{k-q}%
}^{++,u}|X_{-\mathbf{k}}^{v,++}\rangle \!\rangle _E,
\end{eqnarray}
\begin{eqnarray}
(E-t_{\mathbf{k-q+}}-\omega _{\mathbf{q}})L_{\mathbf{kq}}(E) &=&\sqrt{\frac{%
2S}{2S+1}}t_{\mathbf{k-q+}}\langle \!\langle X_{\mathbf{k}}^{++,v}|X_{-%
\mathbf{k}}^{v,++}\rangle \!\rangle _E  \nonumber \\
&&\ \ \ \ \ \ \ \ \ \ -\sum_{\mathbf{p}}t_{\mathbf{k-p+}}(1-n_{\mathbf{k-p}%
})L_{\mathbf{kp}}(E)
\end{eqnarray}
where $n_{\mathbf{k}}=$ $\langle X_{-\mathbf{k}}^{u,++}X_{\mathbf{k}%
}^{++,u}\rangle \!=f(t_{\mathbf{k}+}),\,\omega _{\mathbf{q}}=\omega _{%
\mathbf{q}++}^{+-}.$ Solving this system we obtain the expression which also
has a typically ``non-quasiparticle'' form (without quasiparticle poles, at
least, at not too large current carrier concentration)
\begin{equation}
\langle \!\langle X_{\mathbf{k}}^{++,v}|X_{-\mathbf{k}}^{v,++}\rangle
\!\rangle _E=\frac{2S+1}{E-t_{\mathbf{k}+}+2S/\widetilde{R}_{\mathbf{k}}(E)}
\label{vd}
\end{equation}
\begin{equation}
\widetilde{R}_{\mathbf{k}}(E)=\sum_{\mathbf{q}}\frac{1-n_{\mathbf{k-q}}}{%
E-t_{\mathbf{k-q+}}-\omega _{\mathbf{q}}}.  \label{edw1}
\end{equation}
Note that the result (\ref{vd}) is rigorous for the empty conduction band
(spin-polaron problem) and is in agreement with (\ref{g0})-(\ref{g2}) at $%
I\rightarrow +\infty $. As follows from the spectral representation, $n_{%
\mathbf{k}}^{**}\equiv \langle X_{-\mathbf{k}}^{v,++}X_{\mathbf{k}%
}^{++,v}\rangle \!=0$ (provided that the Green's function
(\ref{vd}) has no poles) since
$\mathrm{Im}\widetilde{R}_{\mathbf{k}}(E<E_F)=0$ because of the
factor $1-n_{\mathbf{k-q}}$. Physically, there is no spin-flip
instability in the $s-d$ model with large $I>0$ since spin-down
electrons are not present in the ground state. The corresponding
spectrum picture is shown in Fig.4

Note that the Green's function (\ref{vd}) itself does not determine an
instability of the saturated state since the spectral representation gives
\begin{equation}
-\frac 1\pi \int dE[1-f(E)]\mathrm{Im}\sum_{\mathbf{k}}\langle \!\langle X_{%
\mathbf{k}}^{\sigma m,v}|X_{-\mathbf{k}}^{v,\sigma m}\rangle \!\rangle
_E=N_{\sigma m}.
\end{equation}
To treat the instability we have to consider the Green's function with $%
\sigma =-$,
\begin{equation}
\langle \!\langle X_{\mathbf{k}}^{-m,v}|X_{-\mathbf{k}}^{v,-m}\rangle
\!\rangle _E=\sum_{\mathbf{q}}\langle \!\langle X_{\mathbf{q}}^{-m,++}X_{%
\mathbf{k-q}}^{++,v}|X_{-\mathbf{k}}^{v,-m}\rangle \!\rangle _E,
\end{equation}
which is proportional to the distribution function $n_{\mathbf{k}}^{**}\!$
and is expressed in terms of the resolvent
\begin{equation}
R_{\mathbf{k}+m}^{**}=\sum_{\mathbf{q}}\frac{n_{\mathbf{k-q}}^{**}}{E-t_{%
\mathbf{k-q}}^{**}+\omega _{\mathbf{q}}}
\end{equation}
where $t_{\mathbf{k}}^{**}$ is the spectrum of new quasiparticles which can
occur with increasing electron concentration.

The orbital instability is determined by the first term in the Hamiltonian (%
\ref{exp2}),
\begin{equation}
\langle \!\langle X_{\mathbf{k}}^{+m,u}|X_{-\mathbf{k}}^{u,+m}\rangle
\!\rangle _E=\frac 1{E-t_{\mathbf{k}m}+(1-c)/R_{\mathbf{k}+m}(E)},
\end{equation}
and is again $S$-independent.

\section{Conclusions}

In this paper we have presented the picture of excitation spectrum in a
saturated conducting (half-metallic) ferromagnet with orbital degrees of
freedom within the framework of simple many-electron models. It is proven
that the half-metallic ferromagnetic state does exist in the presence of
orbital degeneracy. We have focused our consideration on strongly correlated
systems with pronounced term (multiplet) effects. In such a situation the
description that uses atomic many-electron quantum numbers is most adequate;
formally it is provided by the Green's function method for the Hubbard $X$%
-operators.

In contrast with usual itinerant-electron ferromagnets, additional
collective excitation branches (orbitons) occur. Due to the smallness of the
current carrier concentrations, these modes possess low frequencies in
comparison with typical electron energies such as the crystal-field
splitting. Typically, the orbiton energies are of the same order of
magnitude as the magnon energies. Also, mixed excitations with the
simultaneous change of spin and orbital projections exist (``optical
magnons''). All these excitations can be well defined in the whole Brillouin
zone, the damping due to the interaction with current carriers being small
enough.

We have calculated one-particle Green's functions in the non-degenerate
saturated ferromagnetic state. The expressions obtained yield different
criteria for spin and orbital instabilities. It turns out that the saturated
spin ferromagnetism is more stable than the orbital one in the realistic
case $S>1/2$ (e.g., for magnetite and for colossal magnetoresistance
manganites). This means that the half-metallic ferromagnetic phases both
with saturated and non-saturated orbital moments can arise. A more detailed
investigation with the use of real band and spectrum is required for
concrete cases where orbital degrees of freedom yield important
contributions to magnetic properties. In some situations (e.g., near the
quantum phase transitions like those in virtual ferroelectrics \cite{blinc})
the orbital modes can become soft, which should result in orbital
instabilities at smaller conduction electron concentrations $c$.

Due to the electron-orbiton interaction, additional contributions
to the non-quasiparticle density of states can arise. Since the
magnons are gapless, the electron-magnon contribution leads to a
rather sharp one-sided increase of the density of states (with a
crossover energy scale of the order of a typical magnon energy
$\overline{\omega }$) starting just from the Fermi energy, $\delta
g(E)\propto (|E_F-E|/\overline{\omega })^{3/2}$
\cite{UFN,jpc85,ikjpc90}. In the case of half-metallic
ferromagnets with unquenched orbital moment one can predict
additional contributions to $g(E)$ which correspond to orbital
channels. Probably, the most simple way to probe the density of
non-quasiparticle states is to measure the spin-polarized
tunneling current \cite{AI1,ourtransport,falko}. Therefore it
would be very interesting to investigate the density of states of
``suspicious'' materials by, e.g., spin-polarized scanning
tunneling microscopy (STM) technique \cite {STM}, the
non-quasiparticle states being observable below the Fermi energy
for the case of magnetite and above it for the case of colossal
magnetoresistance manganites. It is worthwhile to mention that the
non-quasiparticle states should exist also at the surface of the
half-metallic ferromagnets \cite{KE} since STM is a
surface-sensitive method.

The research described was supported in part by Grant No. 747.2003.2 from
the Russian Basic Research Foundation (Support of Scientific Schools), by
the Russian Science Support Foundation and by the Netherlands Organization
for Scientific Research (Grant NWO 047.016.005).

\newpage

\begin{figure}[tbp]
\includegraphics[clip]{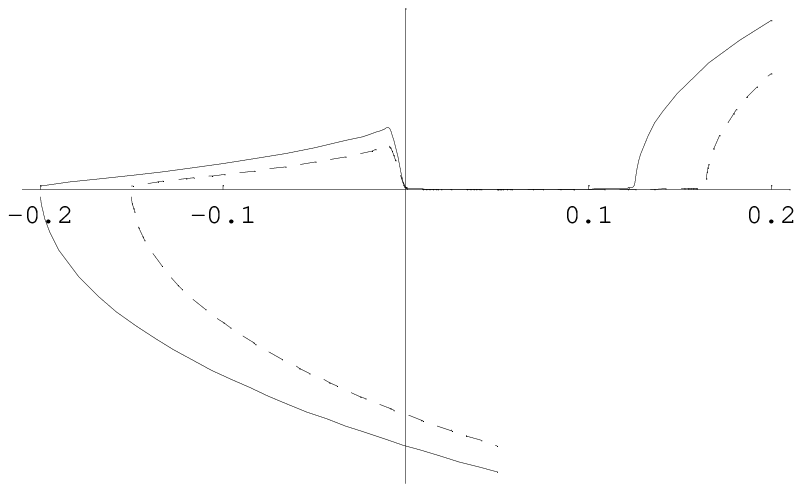}
\caption{Density of states in a half-metallic ferromagnet with double
orbital degeneracy, $I=-0.3<0$ for the semielliptic bare band with the
width of $W=2$, $S=1/2$, crystal-field splitting $\Delta =0.1$.
The energy is referred to the Fermi level.
Solid and dashed lines correspond to crystal-field split subbands, the
corresponding Fermi energies calculated from the band bottoms being
0.15 and 0.2, respectively.
The spin-down subbands (lower half of the figure) nearly coincide with
the bare bands shifted by $IS$.
Non-quasiparticle states in the spin-up subbands (upper half of the
figure) occur below the Fermi level.
}
\label{fig:1}
\end{figure}

\begin{figure}[tbp]
\includegraphics[clip]{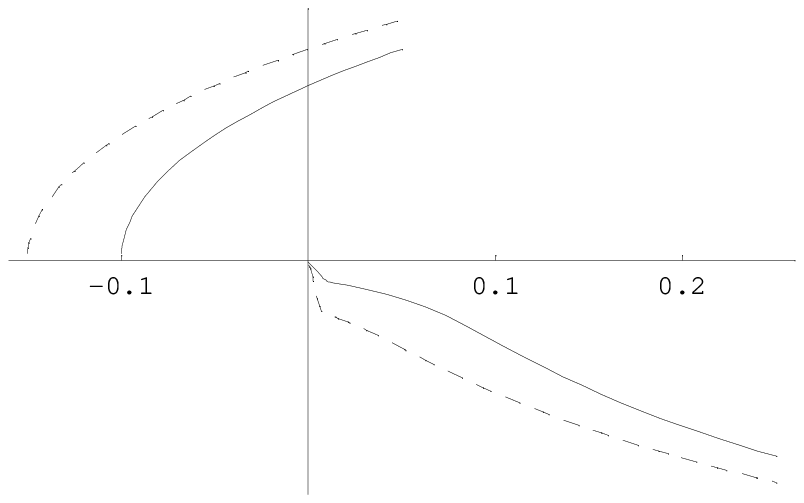}
\caption{
Density of states in a half-metallic ferromagnet with $I=0.3>0$.
The Fermi energies calculated from the orbital subband bottoms are
0.15 and 0.1. Other parameters are the same as in Fig.1
The spin-up subbands (upper half of the figure) coincide with
the bare bands shifted by $-IS$.
The non-quasiparticle tails of the spin-down subbands (lower half of the
figure) occur above the Fermi level.
}
\label{fig:2}
\end{figure}

\begin{figure}[tbp]
\includegraphics[clip]{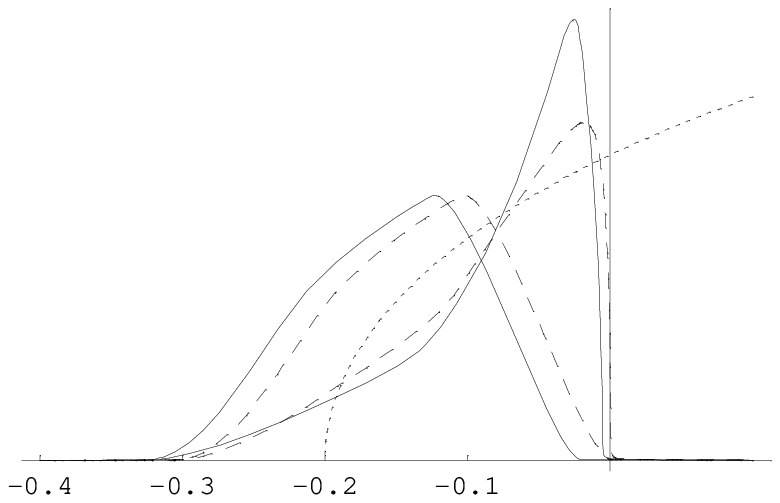}
\caption{ Density of states $g_{\sigma m}(E)$ in a half-metallic
ferromagnet in the model (\ref{HHM}). The Fermi energy calculated
from the bare band bottom is 0.2 ($c=0.034$). Short-dashed line is
the bare semielliptic density of states ($\sigma = +, m=+$). The
solid and dashed lines with peaks near the Fermi level show the
contributions from orbital ($\sigma = +$, $m \ne +$, $\Delta =
0.4$) and spin ($\sigma = -, m=+$) channels, respectively. The
lower curves (without peaks) show the corresponding functions
$-(1/\pi) \rm{Im} \overline{R}_{\sigma m}(E)$. } \label{fig:3}
\end{figure}

\begin{figure}[tbp]
\includegraphics[clip]{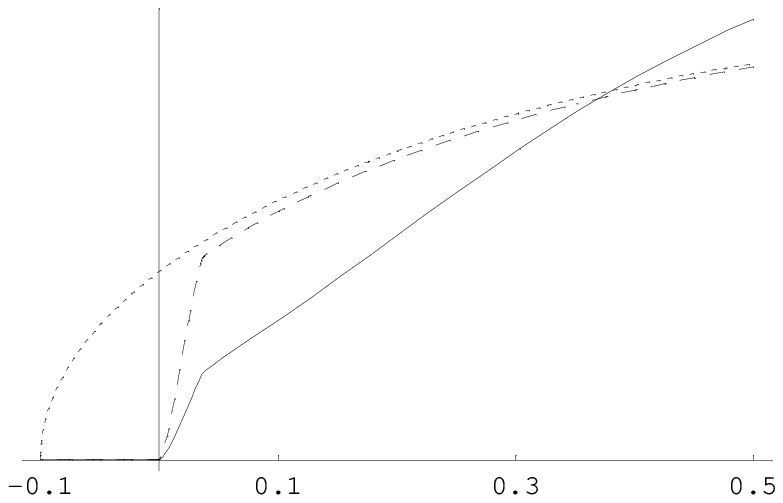}
\caption{
Density of states in a half-metallic ferromagnet in
the model (\ref{exp2}).
The Fermi energy calculated from the bare band bottom is
0.1 ($c=0.019$).
Solid and short-dashed lines correspond to spin down electrons
(the Green's
function (\ref{vd})) and spin up electrons (bare density of states).
The dashed line shows the function
$-(1/\pi) \rm{Im}\overline{\widetilde{R}}_{\sigma m}(E)$.}
\label{fig:4}
\end{figure}

\end{document}